# "Liquid-gas" transition in the supercritical region: Fundamental changes in the particle dynamics

V.V. Brazhkin[1], Yu.D. Fomin[1], A.G. Lyapin[1], V.N. Ryzhov[1], E.N. Tsiok[1] and Kostya Trachenko[2]

[1] *Institute for High Pressure Physics RAS, 142190 Troitsk, Moscow, Russia*

[2] *South East Physics Network and School of Physics, Queen Mary University of London, Mile End Road, London, E1 4NS, UK*

Recently, we have proposed a new dynamic line on the phase diagram in the supercritical region. Crossing this line corresponds to the radical changes of the fluid properties. Here, we focus on the dynamics of model Lennard-Jones and Soft-Sphere fluids. We show that the change of the dynamics from the liquid-like to gas-like can be established on the basis of the velocity autocorrelation function and mean-square displacement. Using the rigorous criterion, we show that the crossover of particle dynamics and key liquid properties occurs at the same line. We further show that positive sound dispersion disappears in the vicinity of this line in both kinds of systems. The dynamic line bears no relationship to the existence of the critical point. We find that the region of existence of liquid-like dynamics narrows with the increase of the exponent of the repulsive part of inter-particle potential.

A liquid near the melting curve has much more in common with a solid than with a gas – this interesting result has been increasingly appreciated recently in the area of liquids, although perhaps not widely by physics community in general. For example, a liquid supports transverse collective modes at high frequency that endow the liquid with shear rigidity at that frequency [1-6]. Many physical properties, including heat capacity, heat conductivity, electroconductivity and so on, change only weakly as a result of melting despite the loss of the long-range order. Close to the melting curve, this behavior is observed at pressure and temperature well above the critical point [5,6]. To denote this state of the liquid, the term "rigid liquid" was proposed [7].

In a "rigid" liquid, particle motion consists of fairly rare jumps over the activation barrier and oscillatory motion in between the jumps. This is reflected in relaxation time $\tau$, the time between two consecutive particle jumps, being larger than the shortest vibration period $\tau_0$ ($\tau_0=2\pi/\omega_0$, where $\omega_0$ is the maximal frequency of transverse modes). This description of liquid dynamics was first proposed by J Frenkel [8], was subsequently re-discovered in a number of papers (see, e.g., Ref. [9]), and was used to calculate thermodynamic and dynamic properties of metallic liquids close to the melting curve [10,11].

Recently, we have shown that the condition $\tau \approx \tau_0$ defines a line, the "Frenkel line" on the pressure-temperature (or density-temperature) phase diagram that separates the state of the "rigid" liquid from the "non-rigid" gas-like liquid [7,12-14]. Crossing the Frenkel line on temperature increase results in the disappearance of shear rigidity at all frequencies, specific heat reaching $2k_B$, particle thermal speed reaching half of the speed of sound and, importantly, the qualitative change of temperature dependence of key system properties such as heat capacity, speed of sound, diffusion coefficient, viscosity and thermal conductivity [7,12,13]. The new line is universal: it separates two liquid states at arbitrarily high pressure and temperature, and exists in systems such as, for example, soft-sphere fluids where liquid-gas transition and the critical point are absent overall. Consequently, the Frenkel line bears no relationship to various versions of "thermodynamic" continuation of the boiling curve such as the so-called "Widom" line [7,12-14].

We proposed [7] that a particularly interesting consequence of crossing the Frenkel line from above is the appearance of "positive sound dispersion" (PSD), the observed increase of the speed of sound at high frequency [4-6]. Recently, the attempt was made to locate the Frenkel line on the basis of molecular dynamics simulations of PSD for supercritical Ar [15].

The condition $\tau \approx \tau_0$ is related to the microscopic dynamics at the Frenkel line, although determining the line on the phase diagram on the basis of this condition can be done in an approximate way only. Indeed, $\tau$ and $\tau_0$ are distributed in a certain range, especially at high temperature. Furthermore, particle motion can be complicated and irregular in the vicinity of the Frenkel line, obfuscating the analysis and separation of the oscillatory and jumping components in particle trajectories. When $\tau$ and $\tau_0$ become comparable, particle motion may not be uniquely separated into oscillations and jumps. Besides, the separation of quasi-harmonic oscillations in a liquid into longitudinal and transverse is not well defined at high temperatures, introducing a uncertainty in $\tau_0$. As a result, the temperature of the Frenkel line defined from the condition $\tau \approx \tau_0$ can be higher than that defined by other physical criteria by a factor 1.4-1.8 [1-3].

The main aim of this work is to identify the dynamic crossover on the basis of uniquely defined physical characteristics: velocity autocorrelation function $Z(t)$ and mean-square displacement $<r^2(t)>$. Interestingly, similar properties were previously explored for supercooled liquids in relation to the problem of liquid-glass transition as well as melting and crystallization processes. On the other hand, no systematic studies of atomic trajectories were done at temperatures considerably exceeding the melting temperature. In this paper, for the first time, we study $Z(t)$ and $<r^2(t)>$ to identify the crossover between the rigid and non-rigid liquid at the Frenkel line. Among other properties, we discuss positive sound dispersion (PSD) and its behavior at the Frenkel line.

The analysis of excitations in a liquid is a complicated task [16]. One of fairly well-developed methods to study liquid dynamics is the instantaneous mode approach [17]. This approach was mostly used to analyze the excitations in a supercooled liquid. One of challenges faced here is that the imaginary part of the vibrational density of states has several contributions that are quite hard to separate (see Ref. [18] and references therein). Furthermore, at high temperature well in excess of melting temperature, when particle dynamics crosses over to the gas-like dynamics, the instantaneous mode approach does not provide an adequate description of excitations in the system.

Velocity autocorrelation function (VAF) $Z(t)$ is defined as

$$Z(t) = <v(0)v(t)> \quad (1)$$

It is well-known that $Z(t)$ for the gas is a monotonically decaying function, whereas for solids and liquids near melting it has both oscillatory and decaying components (see Refs. 16, 19 and references therein). $Z(t)$ for various liquids was studied in detail. In early classic papers [20], it was found that $Z(t)$ for supercritical Lennard-Jones (LJ) fluid may contain both liquid-like and gas-like features, although this fact did not attract subsequent attention. Y. Hiwatari and co-authors have found that depending on the density, $Z(t)$ is qualitatively different in the soft-sphere (SSp) fluids [21]. This was followed by the attempt to relate the oscillations of $Z(t)$ and atomic vibrations in the SSp fluid [22]. More recently, different contributions to $Z(t)$ in a low-viscous liquid with frequent atomic jumps were analyzed in more detail [23], and attempts were made to find distinct features in $Z(t)$ in a supercooled liquid during liquid-glass transition [24]. No detailed analysis of $Z(t)$ was performed with the aim to distinguish the dynamics in a liquid and gas-like fluid.

Similar to $Z(t)$, the time dependence of the mean-square displacement $<r^2(t)>$ is qualitatively different for a liquid and a solid. Consequently, the behavior of $<r^2(t)>$ at long times is used to calculate the diffusion coefficient in liquids and gases and determine melting and crystallization or vitrification points. At long times, the diffusion coefficient, $D$, is related to the mean-square displacement as $D=<r^2(t)>/6t$. Evidently, the time dependence of $<r^2(t)>$ in liquids and gases should be different at short times too, yet no detailed analysis of $<r^2(t)>$ was performed in this time regime.

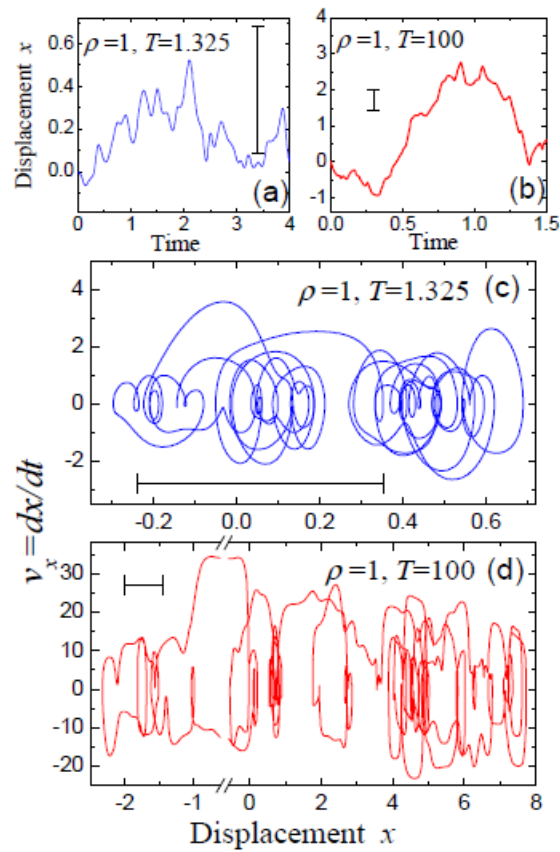

Fig. 1. (colored online) Examples of particle trajectories [(a) and (b)] and phase diagrams (velocity vs displacement) of the particle motion [(c) and (d)] along $x$ axes for the Lennard-Jones (LJ) liquid at different conditions, presented in standard LJ units. Cases (a) and (c) correspond to the "rigid" liquid state with dominance of vibrations. Cases (b) and (d) correspond to collisional motion of the particles in the "non-rigid" gas-like state. Vertical [(a) and (b)] and horizontal [(c) and (d)] bars show average distances between nearest neighbors. Axes break in panel (d) aims to enhance the fragment with intensive collisional motion in the left part.

We note that the second time derivative of $<r^2(t)>$ is proportional to $Z(t)$. Indeed, using the well-known equation (see, for example, Ref. [16]):

$$<r^2(t)> = 6\int_0^t (t-s)Z(s)ds,$$

one has

$$\frac{\partial^2}{\partial t^2}<r^2(t)> = 6Z(t) \qquad (2)$$

Therefore, the analysis of the second derivative of $<r^2(t)>$ and $Z(t)$ is formally equivalent.

Typical trajectories for the LJ liquid at various temperatures and densities corresponding to "rigid" and "non-rigid" liquids are shown in Fig. 1(a,b). We observe a qualitative difference of the type of trajectories in the rigid liquid close to the melting curve and in the non-rigid gas-like fluid. This radical difference is seen in the (displacement, velocity) diagram particularly well (Fig.1(c,d)). At low temperature, we observe the quasi-harmonic motion with relatively rare jumps, whereas at high temperature we see the ballistic collisional regime. The same behavior of atomic trajectories is seen in SSp liquids with different $n$.

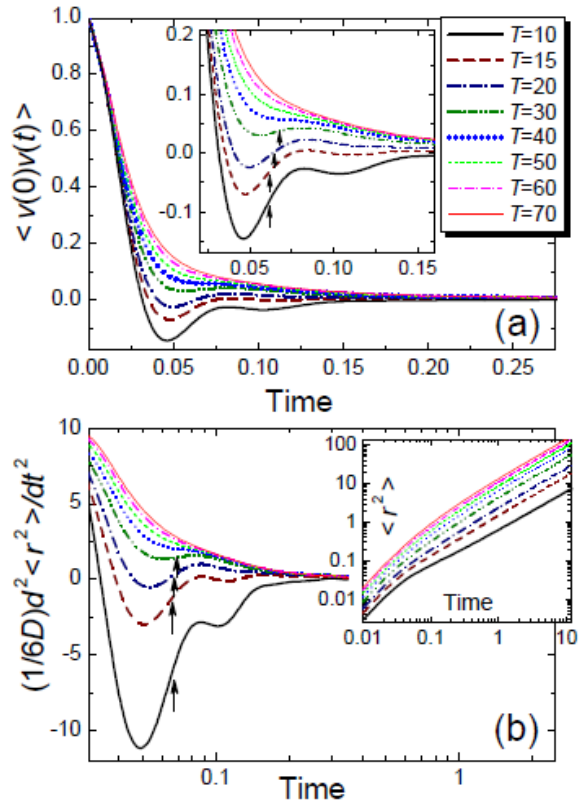

Fig. 2. (colored online) Time dependences of VAF $Z(t)$ (a) and normalized (by $6D$) second derivative of the mean-square displacement (b) for the LJ liquid at pressure $P=200$ and different temperatures. Inset in panel (a) details the main figures; inset in panel (b) shows time dependences of mean-square displacement. Line notations are the same in all figures, and vertical arrows indicate inflection points for the corresponding curves.

We have analyzed the evolution of $Z(t)$ and $<r^2(t)>$ with temperature along several isochors and isobars. In Fig. 2 we show $Z(t)$, $<r^2(t)>$ and its second derivative at several temperature at $P=200$. We observe the qualitative change of the curves with temperature. As discussed below, this change corresponds to the dynamic crossover at the Frenkel line. As a mathematically rigorous criterion of the change of dynamics at the Frenkel line, we propose to demarcate two dynamic regimes by the presence or absence of the positive derivative anywhere at the curve. Indeed, the presence of both decreasing and increasing parts of $Z(t)$ implies that the projection of velocity on the direction of motion changes its sign, i.e. signifies the presence of an oscillatory component of motion. On the other hand, the presence of only decaying $Z(t)$ is implies that a majority of the particles do not have an oscillatory component at short wavelengths. Consequently, the temperature at the Frenkel line can be naturally chosen as the "critical" temperature at which the inflection points (see arrows in Fig. 2) start to coincide with minima and maxima of the function in question. We observe that the second derivatives of $<r^2(t)>$ have the same functional time dependence as $Z(t)$, consistent with Eq. (2). The coincidence of $d^2/dt^2(<r^2(t)>)$ and $Z(t)$ and their temperature evolution serves as a self-consistency check in our calculations.

The line corresponding to the change of dynamics according to the above criterion is shown in Fig. 3. In Fig. 3a we observe that the temperature of the calculated line is about 1.4-1.8 times lower than the line approximately defined from one of the previous criterion, $\tau \approx \tau_0$. The difference is probably related to the uncertainty in determining high-frequency longitudinal and transverse oscillations on the basis of visual analysis of trajectories. Importantly, we observe in Figures 3b,c that the calculated line agrees well (within 10 - 20% difference) with other experimental and theoretical criteria of the Frenkel line proposed previously [7]. We therefore find that the (pressure, temperature) and (density, temperature) phase diagram of the LJ system contains sharp boundaries separating the states of the "rigid" liquid and "non-rigid" gas-like state. For the LJ system at high pressure, the temperature of the Frenkel line exceeds the melting temperature by a factor of 4-5 on the isobar.

The same criterion for the dynamic crossover, based on the qualitative change of $Z(t)$, was used to analyze the SSp liquid with different exponents $n$. Examples of temperature dependence of $Z(t)$ for $n=12$ and $n=36$ are shown in Fig. 4. We find the same qualitative changes for this system as for the LJ liquid seen previously.

For both LJ and SSp systems, we study the positive sound dispersion (PSD) by calculating the dispersion of the longitudinal collective modes $\omega_L(q)$. We detect PSD in the LJ system at low temperature, and find that PSD disappears on temperature increase quite close to the calculated Frenkel line (see Figures 3c, 7b). Importantly, we also find that PSD exists in the SSp system with $n=12$ (see Fig. 5), as it does in the LJ system. Similarly to the LJ system, PSD disappears at high temperature above the Frenkel line. The existence of PSD in the SSp system at low temperature and its disappearance at high temperature is a new result. In particular, this result shows that the presence or absence of PSD is not related to the boiling curve and the critical point (these do not exist in the SSp system) whatsoever, contrary to the previous discussions [6, 15]. Rather, PSD and its crossover are purely dynamic in origin, a point to which we return below.

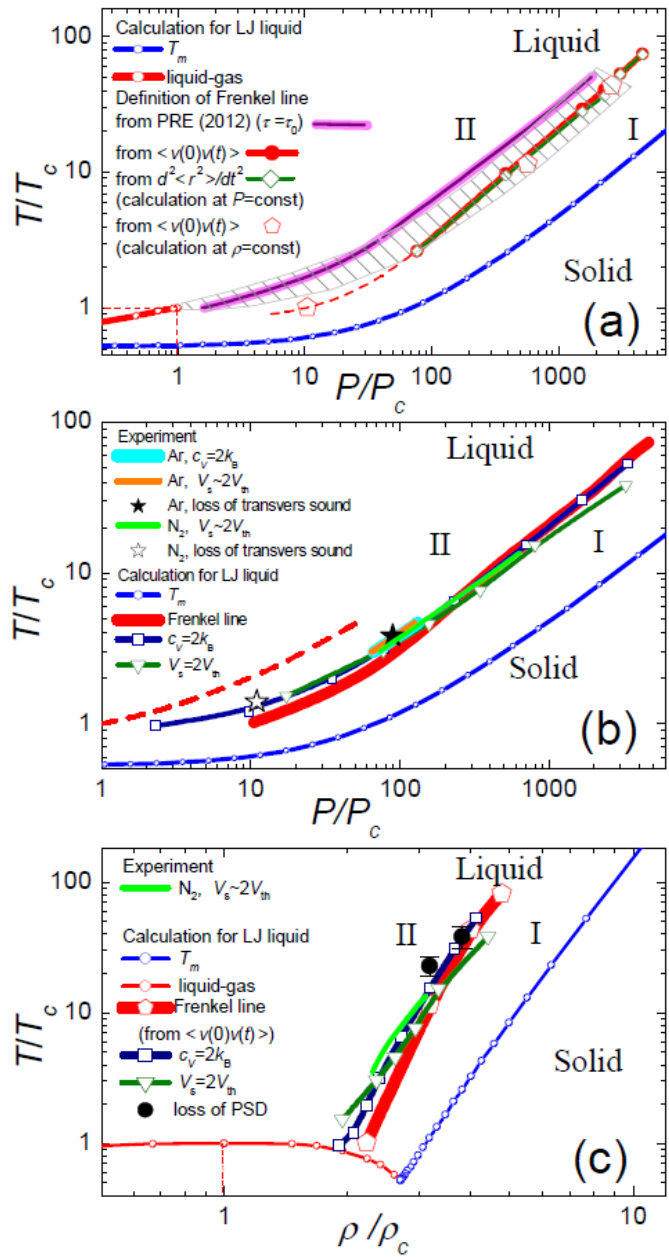

Fig. 3. (colored online) $(P,T)$ phase diagram [(a) and (b)] and $(\rho,T)$ phase diagram (c) of the LJ liquid in the relative critical coordinates. Panel (a) compares previous criteria of the Frenkel line from Ref. [7]. The shaded region covers all experimental and calculated curves from Ref. [7], together with the Frenkel line calculated in this work. Panels (b) and (c) compare positions of the Frenkel line found in this work with calculated and experimental (Ar and $N_2$) curves and points defined by different criteria (see Ref. [7] and references therein). Red dashed line in panel (b) shows the dynamic line proposed in Ref. [15]. In all cases number I correspond to "rigid" liquid state and II – to "non-rigid" gas-like state.

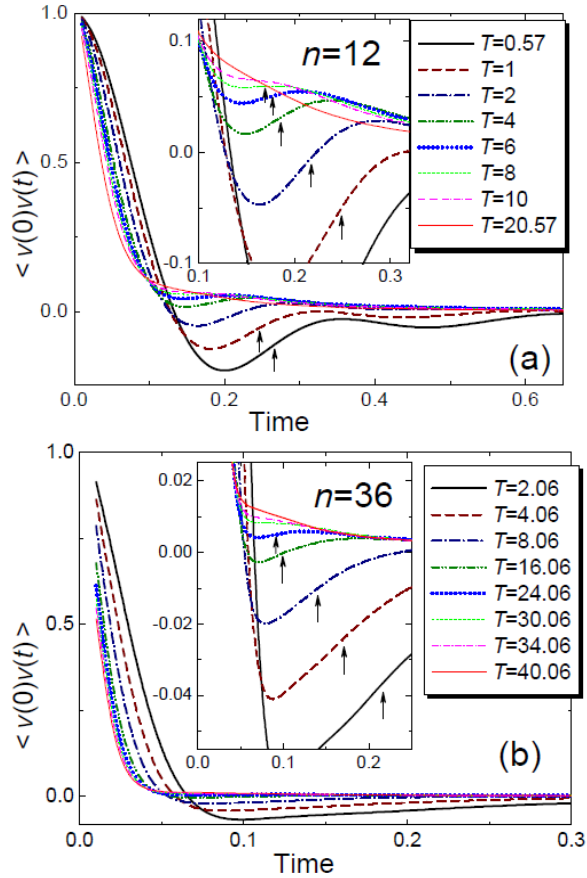

Fig. 4. (colored online) Time dependences of $Z(t)$ for the simulated soft-sphere systems with $n=12$ (a) and $n=36$ (b) with density $\rho =1$ at different temperatures. Insets show a detailed picture of the main figures. Vertical arrows indicate inflection points.

In Figures 6 and 7 we show the calculated Frenkel line on $(P,T)$ и $(\rho,T)$ phase diagrams for the SSp system with $n=6$, 12 and 36. For $n=6$ and 12, the calculated lines slightly differ (within 20%-40%) from the line obtained earlier on the approximate relation $\tau \approx \tau_0$ [7,12,13]. Importantly, we observe that the dynamic crossover line calculated on the basis of $Z(t)$ perfectly coincides with the line that corresponds to $c_v=2k_B$, one of the main criteria of the Frenkel line [7]. Indeed, as discussed earlier [25,26], liquid specific heat reduces from $3k_B$ to $2k_B$ when $\tau \approx \tau_0$, corresponding to the potential energy of shear modes becoming zero due to the loss of shear modes at all frequencies. The perfect coincidence of the crossover lines calculated on the basis of $Z(t)$ and $c_v=2k_B$ is encouraging, and serves as an important self-consistency check of our theory of the dynamic crossover at the Frenkel line. For $n=36$, the agreement between the two lines is not as precise but is satisfactory nevertheless. We also note that the calculated line corresponds to the appearance of PSD at low temperature (see Fig. 7b).

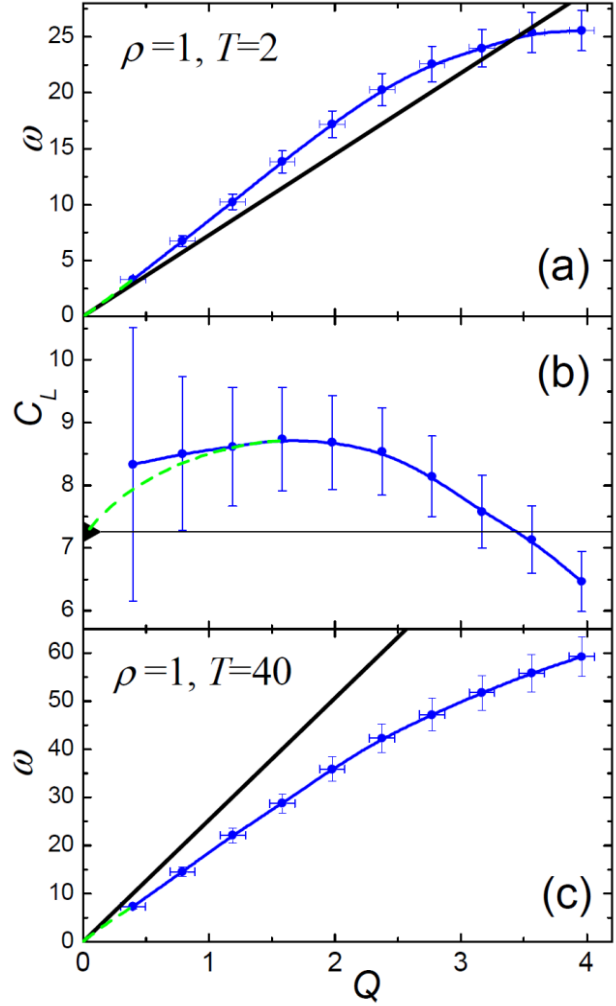

Fig. 5. (colored online) Dispersion curves $\omega(Q)$ (blue circles with bars) for the simulated soft-sphere systems with $n=12$ at low (a) and high (c) temperatures. Panel (b) shows the longitude velocity $C_L=\omega/Q$ vs wave vector $Q$ dependence for the case (a), illustrating evident positive sound dispersion. Thick solid lines in panels (a) and (c) and triangle in (b) correspond to the adiabatic sound velocity at $Q=0$. Dashed lines are extrapolations of the functions to the values at $Q=0$.

We note that for the SSp system, the Frenkel line is parallel in logarithmic scale to the melting line in the whole range of $(P,T)$ parameters whereas for LJ system, the Frenkel line becomes almost parallel to the melting line in the high-pressure range.

Interestingly, the increase of the exponent $n$ results in the fast narrowing of the $P,T$- region where the quasi-harmonic "rigid" liquid exists. For the SSp system, the ratio of the temperature at the Frenkel line to that at the melting line on the isobar is 11.2 for $n=6$; 5.3 for $n=12$ and 2.2 for $n=36$ (see Fig. 6). This behavior is consistent with our earlier findings [7,12] that for $n\approx50$- 60, the line corresponding to $c_v=2k_B$ moves under the melting line, i.e. the region of existence of quasi-harmonic rigid liquid above the melting line disappears. It should be mentioned that for large values of exponents governing the repulsion, $n$, the criterion of the dynamic crossover based on $Z(t)$ no longer applies. Indeed, the oscillatory component of $Z(t)$ becomes hardly identifiable for $n>30$ (see Fig. 4b). Furthermore, the oscillation amplitude becomes extremely small for $n>50$, and the oscillations become irregular themselves. Physically, this means that particles of the fluid spend most of their time outside the field of action of the potential, and move ballistically as in a gas. This is also the reason for a slight discrepancy of the Frenkel line calculated from $Z(t)$ and $c_v=2k_B$ for $n=36$ (see Figs. 6c,7c).

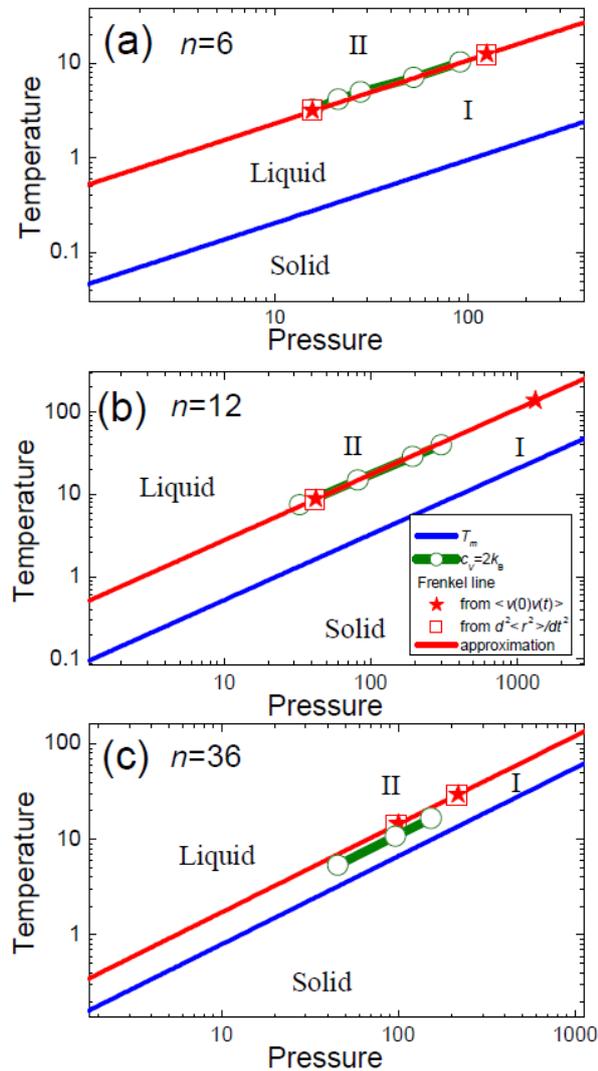

Fig. 6. (colored online) ($P,T$) phase diagrams of the simulated soft-sphere systems with $n=6$ (a), $n=12$ (b) and $n=36$ (c), comparing the location of the Frenkel lines with the lines calculated from the criteria $c_V=2k_B$. In all cases number I correspond to "rigid" liquid state and II – to "non-rigid" gas-like state.

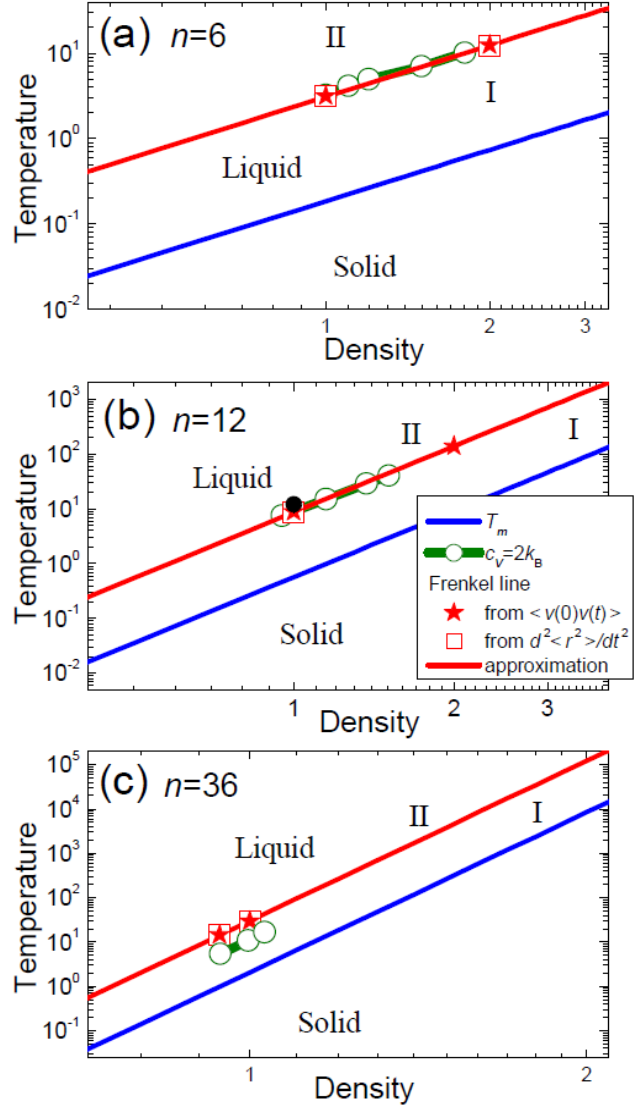

Fig.7. (colored online) $(\rho,T)$ phase diagrams of the simulated soft-sphere systems with $n=6$ (a), $n=12$ (b), and $n=36$ (c). All notations are the same as in Fig.6. In panel (b), solid circle at $\rho=1$ corresponds to the loss of PSD with temperature increase.

As discussed earlier [7,12-14], the Frenkel line and related physics bear no relationship to various versions of "thermodynamic" continuation of the boiling line. This is particularly apparent in our finding that the dynamic crossover at the Frenkel line, including the crossover of PSD, exists in the SSp system where the liquid-gas transition and the critical point are absent altogether. Furthermore, as is seen in Fig. 3b,c, the Frenkel line for the LJ system lies in the range of temperatures that are lower than critical and density that is higher than critical, and starts from the boiling region at temperature $T \approx 0.7\text{-}0.8 T_c$. Thus close to the boiling curve at temperature higher than $0.8T_c$, the dynamics of particles is gas-like on both sides of this curve. In relation to this, we note that at temperature higher than $0.75\text{-}0.85 T_c$, a many-particle system does not possess a cohesive state (i.e. the dependence of the free energy on volume has no minimum), implying that the upper region of the boiling curve should be rather considered as the gas-gas transition (see, e.g. Ref. [27]). The close proximity of the temperature of the dynamic crossover and the disappearance of the cohesive condensed state is obviously not coincidental.

As mentioned above, there has been a recent attempt [15] to determine the dynamic Frenkel line for supercritical Ar on the basis of molecular dynamics simulations of the PSD (it is shown in Fig. 3b as a dashed line). We observe that the line proposed in Ref. [15] is notably different from the Frenkel line calculated in our work. One can mention that PSD is not a fundamental phenomenon on its own: as was discussed by J Frenkel long time ago (see section IV of Ref. 8), the existence of PSD is simply the result of the presence of high-frequency transverse-like collective excitations. These excitations disappear at the dynamic crossover at the Frenkel line as discussed above. The existence or disappearance of PSD is therefore only one of many consequences of the dynamic crossover rather than its origin. As shown above, PSD exists in the SSp system where no boiling line and critical point exist. This additionally points to the conceptual inconsistency, from the physical point of view, of using the combination of thermodynamic continuation of the boiling line on one hand and emergence of PSD on the other hand, to study the dynamic crossover as proposed in Ref. 15. We finally note that the observation of PSD in liquids by inelastic x-ray scattering [4-6] is certainly important from the experimental point of view. At the same time, quantifying PSD in molecular dynamics simulations suffers from quite large uncertainty and errors, and can not be, in our view, used as a convenient computation criterion to establish the location of the dynamic line. It suffices to note that calculated parameters at which PSD disappears for Ar, calculated by the same authors [6,15], differ widely. Indeed, at $T=3.8T_c$, PSD disappears at $P<80P_c$ according to Ref. 6, whereas PSD disappears at $P<30P_c$ according to Ref. 15.

In summary, we proposed a criterion to locate the dynamic line on the phase diagram of fluids that is both mathematically rigorous and convenient for simulations. The calculated Frenkel line for the LJ and SSp systems coincides with various experimental data for rare-gas fluids as well as the data calculated on the basis of other important physical criteria such as $c_v=2k_B$. For the LJ system, the Frenkel line lies in the range of temperatures lower than critical and densities higher than critical, and starts from the boiling curve at $T≈0.7$-$0.8T_c$. The region of "rigid" liquid narrows with the increase of the exponent of the repulsive part of inter-particle potential, and disappears when the exponent is in excess of about 50. The new criterion opens an exciting possibility to calculate and map the dynamical line for various liquids with different types of structure and interactions.

**Methods**

We have studied Lennard-Jones (LJ) liquid and Soft Spheres (SSp) liquids with different exponents $n$ and in a very wide range of parameters. An essential property of SSp system is that the phase diagram corresponds to the equation $\gamma = \rho\sigma^3 \left(\dfrac{\varepsilon}{k_B T}\right)^{3/n} = $ const. Here, $\gamma = 2.33$ for $n=6$; $\gamma = 1.15$ for $n=12$ and $\gamma = 0.942$ for $n=36$ [28]. In the simulations of LJ liquid, system size varied depending on the density reaching 4000 particles at the highest density. The equations of state were integrated using velocity Verlet algorithm. The temperature was kept constant during the equilibration by velocities rescaling. When the equilibrium was reached, the system was simulated in NVE ensemble. The usual equilibration and production runs consisted of 1.5 million and 0.5 million steps, respectively with the time step of 0.001 in LJ units. The SSp system was simulated in NVE ensemble. The system consisted of 1000 particles, with the time step of 0.0005 in reduced units. The equilibration and production runs consisted of 3.5 million and 0.5 million steps, respectively. The simulations and computation of properties were made in the same way as for the LJ system. In parts of the text and figures, densities, temperatures and pressures of the LJ liquid are given in the reduced units $\rho/\rho_c$, $T/T_c$, and $P/P_c$. The following critical parameters, averaged from literature sources, were used for the LJ system: $\rho_c=0.314$, $T_c=1.31$, $P_c=0.13$. The temperature corresponding to $c_V = 2.0$ was determined from the dependence of the isochoric heat capacity on temperature along the isochors by linear interpolation of the data. The dispersion of the longitudinal collective excitations $\omega_L(q)$ has been calculated for both LJ and SSp systems using the same approach as in Ref. [15]. The dispersion has been analyzed at different temperatures along the several isochors: $\rho=1$ for the SSp (n=12) system and $\rho=1, 1.2$ for the LJ system.

Acknowledgments:
The authors wish to thank S.M. Stishov and N.M. Chtchelkatchev for valuable discussions. The work has been supported by the RFBR (11-02-00303, 11-02-00341 and 13-02-01207) and by the Programs of the Presidium of RAS. K.T. is grateful to EPSRC.